\documentstyle[epsfig,12pt]{article}
\begin{document}

\title{$\pi\pi$ S-WAVE INTERACTION AND $0^{++}$ PARTICLES
\thanks{Talk presented at 34th Course of International School of
Subnuclear Physics, Erice, Sicily, July 3-12,1996}
}

\author{ Bing-Song Zou \\
Queen Mary and Westfield College\\
London E1 4NS, United Kingdom}
\date{}

\maketitle
\begin{abstract}
From a re-analysis of $\pi^+\pi^-\to\pi^+\pi^-$ and $\pi^+\pi^-\to K\bar K$
data, we found for $\pi\pi$ S-wave interaction below 2 GeV, $\sigma(400)$,
$f_0(980)$, $f_0(1500)$ and $f_0(1780)$ clearly show up. $f_0(1370)$ can be
included with a very small branching ratio to $\pi\pi$. $f_0(1300)$
and $f_0(1590)$ are not real resonances and are due to
interference effects of above resonances. The $\sigma(400)$ with a width
about 700 MeV is mainly produced by t-channel exchange force. 
\end{abstract}


Nearly all known mesons are bound states of a quark and an antiquark.
The $q\bar q$ mesons containing u, d and s quarks can be ascribed into 
various $^{2S+1}L_J$ SU(3) nonets according to their spin S, orbital angular 
momentum L and total angular momentum J. Among the lowest nonets,
$^1S_0$, $^3S_0$ and $^3P_2$ nonets are well established; $^3P_1$ and
$^1P_1$ are also settled although there are still some
uncertainties \cite{PDG}; only $^3P_0$ nonet is very problematic:
it has two opening positions for its isoscalar part, while
there are too many candidates, but none of them can fill in 
without controversy. Now let's have a look at these candidates.

\begin{itemize}
\item $\sigma(300\sim 800)$ with width 
$\Gamma = 200\sim 1000 MeV$\cite{PDG1}.

It was listed in old PDG booklets\cite{PDG1} more than twenty years ago,
but has been dropped out in newer versions. However it is needed in the 
$\sigma$ model, the extended Nambu-Jona-Lasinio model and the 
nucleon-nucleon scattering models; it is also needed to explain
the low energy enhancement in $\pi\pi$ invariant mass spectra from
various production processes.

\item $f_0(980)$ with a peak width about 50 MeV\cite{PDG}.

Due to its very narrow peak width, it was regarded to be difficult
to be ascribed as a $q\bar q$ state. Many exotic explanations have
been proposed for it, such as $K\bar K$ molecule\cite{Isgur, Julich},
multiquark state\cite{Jaffe}, and Gribov's minion \cite{Gribov}, etc.

\item $f_0(1300)$ with $\Gamma = 200\sim 400 MeV$ and 
$\Gamma_{\pi\pi}/\Gamma > 90\%$.

It was commonly ascribed as a $q\bar q$\cite{PDG}. But we will see that
it in fact does not exist.

\item $f_0(1370)$ with $\Gamma = 200\sim 400 MeV$ and 
$\Gamma_{\pi\pi}/\Gamma < 20\%$.    

It is needed in fitting Crystal Barrel data on $\bar pp$ annihilations
\cite{CB1,CB2} and some other data\cite{PDG}, and was found decaying into
$\pi\pi$, $\eta\eta$ and dominantly to $4\pi$. But it needs further
confirmation\cite{PDG}.

\item $f_0(1500)$ with $\Gamma = 90\sim 150 MeV$.

This resonance was observed to decay into $\pi^0\pi^0$, $\eta\eta$, 
$\eta\eta'$ and $4\pi^0$ in $\bar pp$ annihilations by Crystal Barrel
Collaboration\cite{CB1,CB2}. Then clear signals for it were also found
in $J/\Psi\to\gamma 2\pi^+2\pi^-$\cite{Bugg} and central production processes
of $pp\to pp\pi^+\pi^-$ and $pp\to pp(2\pi^+2\pi^-)$\cite{WA}.
All these production processes are traditionally believed to favour 
glueballs and the mass of the $f_0(1500)$ is very close to the lattice-QCD
prediction by UKQCD group\cite{UKQCD}. Therefore the glueball explanation
was suggested for it\cite{Close}. Meanwhile it was also suggested to be
$\rho\rho -\omega\omega$ molecule\cite{Torn1} and $q\bar q$ state\cite{Zou}.

\item $f_0(1590)$ with $\Gamma = 160\sim 200 MeV$.

The $f_0(1590)$ was observed by GAMS collaboration in $\pi^-p$ reactions
at 38 GeV/c and was regarded as a glueball candidate for a long time
\cite{PDG}.

\item $f_J(1710)$ with $\Gamma\approx 140 MeV$.

The $f_J(1710)$ has been clearly seen in ``glue rich" $J/\Psi$ radiative
decay and its spin may be 0. But in central production, a structure at the
same mass was seen, but favors spin 2. So its spin is still 
uncertain\cite{PDG}.

\item $f_0(1750-1820)$ with $\Gamma\approx 150 MeV$.

There are some new evidences for this resonance in $J/\Psi$ radiative
decay\cite{Bugg} and $\gamma\gamma$ fusion\cite{L3}. It is not 
established yet.

\end{itemize}

From the list, there are obviously too many $0^{++}$ particles to be
explained as $q\bar q$ mesons. Even assuming two radial excitation 
$^3P_0$ $q\bar q$ states below 2 GeV, we should only have four $0^{++}$
$q\bar q$ mesons. Then what are the others? Does it mean there are definitely 
some exotic particles among them? Before we make any conclusion we should 
consider another possibility: not all of them are real resonances. 

A good place to examine them is S-wave $\pi\pi$, $K\bar K$ and $\eta\eta$
scattering amplitudes. If a $0^{++}$ resonance has a substantial coupling
to $\pi\pi$, then it should show up clearly in the $\pi\pi\to\pi\pi$ S-wave
amplitude. Therefore we first examined the existing and
commonly used $\pi\pi\to\pi\pi$ S-wave phase shifts\cite{CM}. We found
\cite{ZB1,ZB2}
that among $0^{++}$ particles listed above only a broad $\sigma(400)$
with a width about 700 MeV and the $f_0(980)$ clearly show up.
The broad $\sigma(400)$ can be naturally explained by the t-channel
$\rho$ exchange\cite{Julich,ZB2}. The $f_0(980)$ has a large decay width
about 400 MeV, but appears as a narrow structure with a width
about 46 MeV due to $K\bar K$ threshold effect\cite{ZB1}. It is
dominantly $s\bar s$ mixed with $K\bar K$ virtual states\cite{ZB2,Torn2}.

\begin{figure}[htbp]
\begin{center}\hspace*{-0.cm}
\epsfysize=11.0cm
\epsffile{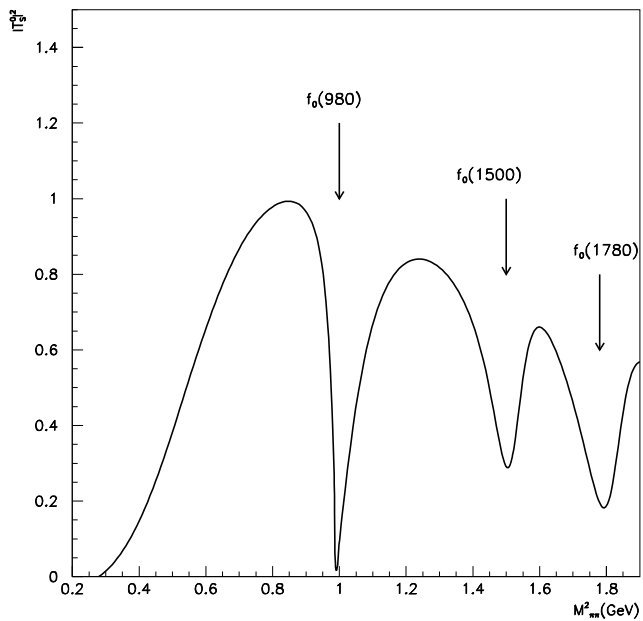}
\end{center}
\caption{
The isoscalar $\pi\pi\to\pi\pi$ S-wave amplitude squared.
}
\label{fig:ampl}
\end{figure}

Then how about other $0^{++}$ particles? From modern
experimental results, the old CERN-Munich solution\cite{CM} of $\pi\pi$
S-wave phase shifts is questionable for energies above 1200 MeV.
As a second step we re-analysed their original data for 
$\pi^-p\to\pi^-\pi^+n$
at 17.2 GeV. We found\cite{BSZ} that $f_0(1500)$ clearly shows up
in the $\pi\pi$ S-wave phase shifts.  Recently we also re-analysed
the original data for $\pi^+\pi^-\to K\bar K$ from Argone\cite{Argone}
and Brookhaven\cite{Brook}. We found\cite{BZ} that a $f_0(1750-1820)$
is needed. The isoscalar $\pi\pi\to\pi\pi$ S-wave amplitude squared is
obtained as shown in Fig.\ref{fig:ampl}. This is a very interesting spectrum.
The broad $\sigma(400)$ by t-channel $\rho$ exchange provides a very broad
background, three resonances $f_0(980)$, $f_0(1500)$ and $f_0(1780)$
superpose on it and therefore appear as dips. The peaks at 800, 1300
and 1590 MeV are caused by the interference effects, so they are not
additional resonances. For $\pi\pi\to\eta\eta$ S-wave intensity,
the peak of $f_0(1590)$ is also caused by a dip around $f_0(1500)$.

As to $f_0(1370)$, since it has a very small branching ratio to $\pi\pi$,
it may have little influence on $\pi\pi$ S-wave amplitude and therefore is not
excluded, but needs further confirmation.
For $f_J(1710)$ it is quite possible containing two components.
Preliminary result from BES Collaboration\cite{BES} suggests that
it is composed of a $2^{++}$ component below 1.7 GeV and a $0^{++}$
resonance at $f_0(1780)$. 

In summary, for $\pi\pi$ S-wave interaction below 2 GeV, $\sigma(400)$,
$f_0(980)$, $f_0(1500)$ and $f_0(1800)$ clearly show up. $f_0(1370)$ can be
included with a very small branching ratio to $\pi\pi$. $f_0(1300)$
and $f_0(1590)$ are not real resonances and are due to
interference effects of above resonances. The $\sigma(400)$ with a width
about 700 MeV is mainly produced by t-channel exchange force. The $f_0(980)$
has a large decay width as well as a narrow structure width due to
$K\bar K$ threshold effect. It is dominantly $s\bar s$ mixed with
$K\bar K$ virtual states, as well as some $u\bar u+d\bar d$. 
Others can also be accommodated by $1 ^3P_0$ and $2 ^3P_0$ $q\bar q$ nonets,
though we cannot exclude the possibility that $f_0(1500)$ may be a
glueball or glueball mixed with $q\bar q$.
All $0^{++}$ $q\bar q$ mesons are expected
to have large mixing of $s\bar s$ with $u\bar u + d\bar d$ \cite{OZI}.
They may also have admixture of virtual meson-meson states and some
glue components. 

\medskip
{\bf Acknowledgement:} 
It is a pleasure to thank Prof. T.H.Ho for his nomination and the
organisers for their invitation to participate this nice school.
I am greatly indebted to Prof. D.V.Bugg for his advice and collaboration.
I also gratefully acknowledge the support of K.C.Wong
Education Foundation, Hong Kong, for a visit to Beijing where part of
this talk was prepared and presented.

\end{document}